\newcount\eqnno\eqnno=0\newcount\secn\secn=0
\def\numeqn{\global\advance\eqnno by 1 \eqno(\the\eqnno)}
\font\bigbf=cmbx10 scaled\magstep2
\font\bigbish=cmbx10 scaled\magstep1
\def\stringbook{1}
\def\witten{2}
\def\acdpp{3}
\def\acdrhb{4}
\def\everett{5}
\def\hillwidrow{6}
\def\vortref{7}

\def\acdwbp{8}
\def\mandm{9}
\def\cinderella{10}
\def\EP{11}
\def\JR{12}
\def \ni{\noindent}
\magnification=\magstep1
\pageno=0\nopagenumbers
\rightline{DAMTP 96-71}
\rightline{SWAT/126}
\bigskip\bigskip
\centerline{\bigbf  GENERIC CURRENT-CARRYING STRINGS}
\vskip 1 true in
\centerline{\bigbish Anne-Christine Davis$^1$}
\centerline{\bigbish Warren B. Perkins$^2$}
\bigskip
\centerline{1.\ \it Department of Applied Mathematics and Theoretical Physics,}
\centerline{\it Silver Street, Cambridge, CB3 9EW, Great Britain}
\centerline{2.\ \it Department of Physics, University of Wales Swansea,}
\centerline{\it Singleton Park, Swansea, SA2 8PP}
\vskip 1 true in
\centerline{\hfill{\bf ABSTRACT}\hfill}\nobreak\smallskip\noindent

We investigate the standard model in a cosmic string background and show that
the electroweak symmetry is partially restored. For a range of parameters
the electroweak Higgs field can wind in this region, producing an electroweak
string stabilised by the cosmic string background. In this case there are
quark and lepton zero modes that result in the string becoming superconducting
at the electroweak scale. If the electroweak Higgs field does not wind, there
are no zero modes, but there are fermion bound states. These bound states 
can also carry a current.
\def\startnumberingat#1{\pageno=#1\footline={\hss\tenrm\folio\hss}}
\vfill\eject
\startnumberingat1
\def\e{{\rm e}}
\noindent{\bf 1) Introduction}
\overfullrule=0pt
\smallskip
Topological defects, and in particular cosmic strings, are likely to arise
as a result of phase transitions in the early universe. A network of such 
strings could explain the observed large scale structure and the anisotropy
in the microwave background radiation [\stringbook]. They arise in many Grand Unified
Theories (GUT) of particle physics and have high predictive power, the 
only free parameter being the mass per unit length. Recently, the possibility
of superconducting strings [\witten], or current carrying strings in general,
has received a revival of interest. This has been sparked by the 
realisation that current carrying strings may be more general than 
previously thought [\acdpp]. The cosmology of such strings could be modified
considerably, see [\acdrhb] and references therein. Hence, it is important to 
ascertain whether or not strings carry a current, and how large the current is.

Cosmic string superconductivity can arise as a result of boson condensates,
fermion zero modes [\witten], or gauge boson condensates [\everett]. Strings can 
also carry a current as a result of fermion bound states [\hillwidrow]. The maximal 
current can be enormous, approaching the grand unified scale for GUT 
strings. Models containing strings carrying such enormous currents can be ruled out as they produce
stable remnants, vortons [\vortref], which would dominate the energy density 
of the universe. However, it was pointed out that strings could become
current carrying at a subsequent phase transition [\acdpp] as a result of the
restoration of, say,  the electroweak symmetry around the string. In this case
the resulting current would be proportional to the electroweak scale,
it would produce only a modest change to the standard string cosmology,
but could still change the microphysics of the string significantly.

In this paper we address this issue. 
We show that couplings  between the string and the electroweak sector lead to 
at least partial  restoration of the
electroweak symmetry around the string.
Such couplings arise either as  direct couplings given by the GUT,
or as effective couplings at higher order. 
 Depending on the ratio 
of the couplings, as given by the GUT, the Higgs field can wind. If it
does wind then there are zero modes of the ordinary quarks and leptons 
which give rise to fermion superconductivity.
If the Higgs field doesn't wind then there are no zero modes, but there
are bound states of the ordinary quarks and leptons. This is because there
is an effective two dimensional potential, which always has at least one 
bound state. These bound states can carry a current [\hillwidrow]. 
Consequently, we conclude that a cosmic string can always carry a 
current, though the current may not be enormous, or persistent.
  
\bigskip
\noindent{\bf 2) The Electroweak $\times U(1)$ Model}
\smallskip
We work with a model in which we augment the Weinberg-Salam model  with a string forming U(1) scalar field $\tilde\chi$
and associated gauge field $X_\mu$ [\acdwbp].
The Lagrangian we consider is
$$
{\cal L}= (\tilde D_\mu \tilde\chi)^\dagger(\tilde D^\mu \tilde\chi)-{1\over 4}X_{\mu\nu}X^{\mu\nu}
-\tilde\lambda(\tilde\chi^\dagger\tilde\chi-\tilde\eta^2)^2 
-\xi M^2(\tilde\chi^\dagger\tilde\chi-\tilde\eta^2)\Phi^\dagger\Phi
$$
$$
+(D_\mu \Phi)^\dagger(D^\mu \Phi)-{1\over 4}Y_{\mu\nu}Y^{\mu\nu}
-{1\over 4}W^a_{\mu\nu}W^{a\mu\nu}-\lambda(\Phi^\dagger\Phi-\eta^2)^2,
\numeqn$$
where the  $\tilde\chi$ covariant derivative is 
$$
\tilde D_\mu =\partial_\mu -iqX_\mu,
\numeqn$$
and the electroweak Higgs field has a charge $\alpha$ under the extra U(1) 
so that its covariant derivative is
$$
D_\mu =\partial_\mu -i\alpha X_\mu-{i\over 2}g {\bf \tau.W}_\mu-{i\over 2}g'Y_\mu.
\numeqn$$ 
We have also introduced a scalar-scalar cross coupling with  magnitude $M^2$  and sign $\xi=\pm 1$.
These couplings between the string and electroweak sectors are expected to arise   generically
either directly from  the underlying GUT or as effective couplings at higher order.

We can write the quartic scalar terms in the Lagrangian in the form,
$$
V_4 =\pmatrix{ \tilde\chi^\dagger\tilde\chi &\Phi^\dagger\Phi \cr}
\pmatrix{\tilde\lambda & \xi M^2/2 \cr
        \xi M^2/2 &  \lambda \cr}
                              \pmatrix{ \tilde\chi^\dagger\tilde\chi \cr
                                                    \Phi^\dagger\Phi \cr}
\numeqn$$
$V_4$ is unbounded below if the matrix has a negative eigenvalue, requiring that
this is not the case yields the constraint 
$$
 M^2/2 <\sqrt{\tilde\lambda\lambda}.
\numeqn$$ 
Thus there is a natural limit to the magnitude of the cross coupling.
 
When the extra U(1) symmetry is broken a string forms with the usual profile 
for the Higgs and gauge fields;
$$
\tilde\chi\sim\e^{i\theta}\cases{ar & small $r$ \cr \tilde\eta & large $r$\cr}
\numeqn$$
$$
X_\theta \sim\cases{br & small $r$ \cr 1/qr & large $r$\cr}
\numeqn$$
with the other components of $X_\mu$ being zero as usual. For concreteness we will call the 
energy scale of the string
the GUT scale, but our analysis does not depend on the energy scale of string formation.
\medskip
\noindent{\bf 3) The Electroweak Background Fields}
\smallskip
We can now consider the electroweak gauge and Higgs field profiles around the
string. Whilst this was done in [\acdwbp], the Higgs-Higgs interaction was 
neglected and a more general analysis is necessary.
We work in temporal gauge and look for static solutions of the equations of motion. 
We can consistently set most of the electroweak fields to zero and look for solutions with only
$Z_\mu$ ($Z_\mu=-{g\over G}W^3_\mu+{g'\over G}Y_\mu$, $G^2=g^2+g'^2$)
and the lower component of the Higgs doublet, $\phi$, non-vanishing.
The Lagrangian then reduces to
$$
{\cal L}=-\phi^2_{,r}-( {G\over 2}Z_\theta +\alpha X_\theta)^2 \phi^2
-{1\over 2}(Z_{\theta,r} +Z_\theta/r)^2-\lambda(\phi^2-\eta^2)^2
 -\xi\tilde M^2(\chi^\dagger\chi-1)\phi^2
\numeqn$$
where  $\tilde\chi=\tilde\eta\chi$ and $M^2\tilde\eta^2=\tilde M^2$.

Applying the scalings: $\phi={\tilde M\over\sqrt{\lambda}}\psi$, $ r=x/\tilde M$ 
and $GZ_\theta/\tilde M=z$ leads to the equations of motion,
$$
{1\over x}{d\over dx}(x\psi_{,x})=\bigl[( {z\over 2} +{\alpha\over
\tilde M} X_\theta)^2-2\eta^2{\lambda\over \tilde M^2}+\xi(\chi^\dagger\chi-1)\bigr] \psi
+2\psi^3
\numeqn$$
$$
{1\over x}{d\over dx}(x(z_{,x}))={z\over x^2}+{G^2\over\lambda}( {z\over 2} 
+{\alpha\over\tilde M} X_\theta)\psi^2 
\numeqn$$

We have solved these equations numerically for a range of parameters. Some typical profiles are presented in figures 1-3. 
The important parameters in the model are the  ratio of the new U(1) charges of the GUT Higgs and electroweak Higgs,
 ${\alpha\over q}$, the Higgs-Higgs cross coupling, $M$, and the ratio of the electroweak energy scale to the GUT scale.
In fig.1 we show the variation with the Higgs-Higgs coupling, M; fig.2
shows the variation with the ratio of the gauge couplings $\alpha/q$;
fig.3 shows the variation with the ratio of the two length scales. 
 The corresponding Z profiles are shown in fig.4. The solid line corresponds to the same parameter values in all four
figures and only the specified parameter is varied in each case.

The gauge coupling gives a positive contribution to the electroweak Higgs field mass term and the Higgs-Higgs
cross coupling gives a contribution whose sign is determined by $\xi$. The latter contribution is restricted to the 
GUT string core, while the former varies as $1/r^2$ outside the core. 
In fig.1 we see the effects of the scalar cross coupling at small distances, inside the GUT string core the scalar
cross coupling increases the Higgs expectation value, while outside
 the Higgs VEV is reduced due to the
interaction with the string's gauge field.
As the GUT length scale gets
smaller  relative to the electroweak scale the effects of the gauge coupling become dominant and the electroweak 
Higgs field is pushed towards zero at the core of the string. If the $1/r$ form of the GUT gauge field were to persist 
into the core of the string, the electroweak Higgs field would be forced exactly to zero at the centre of the string as
the only regular solution of the equation of motion is $\phi\propto r^{\alpha/q}$. This limit is considered in ref.{\mandm}.
The onset of this behaviour is displayed in fig 3.
If we
attempt to construct a power series solution in the region $m_{EW}r < 1$, $m_{GUT}r>1$ the string gauge
field varies as $1/r$ and the Higgs field varies as $r^{\pm \alpha/q}$. As $r_{GUT}$ gets smaller the
electroweak Higgs field at the centre of the string must drop to zero so as not to overshoot its asymptotic value.
Thus as the GUT mass scale gets larger $\psi(0)$ tends to zero. As the $1/r$ form of the gauge field 
does not persist to $r=0$, the Higgs field must have zero gradient at the origin, but the departure from the $r^{ \alpha/q}$ form is restricted to the region $r<r_{GUT}$.

The profiles shown confirm our previous results [\acdwbp] where a region of electroweak symmetry
restoration of order the electroweak scale was found by considering the gauge coupling alone. We have a `composite 
string'; the GUT string core is surrounded by a region of partially restored 
electroweak symmetry containing a Z magnetic flux.

If we introduce a winding into the electroweak Higgs field, i.e. $\phi = \phi \e
^{il\theta}$, then 
$\alpha X_\theta$ is simply
 replaced by $\alpha X_\theta+l/r$. For $r>r_{GUT}$ the ratio of gauge couplings,  ${\alpha \over q}$ is replaced
by ${\alpha \over q} +l$ while as $r\to 0$ we have a persistent $1/r$ behaviour that forces $\psi(0)=0$
exactly. This change of boundary condition has little effect for $m_{GUT}\gg m_{EW}$, as discussed above,
and the winding merely introduces an effective gauge coupling. The energy of the electroweak fields around
the string increases as the gauge coupling ratio, ${\alpha \over q}$, increases as shown in fig.5. For
$g>0.5$ we expect the electroweak field energy to be minimised by introducing a winding in the electroweak Higgs field. This is consistent with our previous results [8].
In this case we really have an electroweak string surrounding the GUT core.
The difference here is that the electroweak string is stabilised by the 
presence of the GUT core. The appearance of such Higgs field windings is considered in ref.{\mandm} in terms
of the 'Cinderella' process seen in the pure electroweak string[\cinderella].
 
\def\e{{\rm e}}
\bigskip
\noindent{\bf 4) Fermionic States in the String Background}

In this section we analyse the Dirac spectrum in the background of our string. We assume that the string has the most
 favourable winding and work in a gauge where there is no W condensate
and the only non-zero electroweak fields are $\phi$ and $Z_\theta$. First we
consider the case where the electroweak Higgs field has a non-trivial winding
and search for fermion zero modes. We consider the non-winding case later in
this section and search for fermion bound states.

In the background  of the composite string the quark Lagrangian is
$$
\eqalign{
&{\cal L}_{\rm  quark}= \cr &
\bar u_L \gamma^\mu(-i\partial_\mu -{1 \over 2G}(g^2-g'^2/3)Z_\mu +\nu X_\mu)u_L +
\bar d_L \gamma^\mu(-i\partial_\mu +{1 \over 2G}(g^2+g'^2/3)Z_\mu+\nu X_\mu)d_L
\cr &
+\bar u_R\gamma^\mu(-i\partial_\mu+{2\over 3}{g'^2\over G} Z_\mu+\nu' X_\mu)u_R
+\bar d_R\gamma^\mu(-i\partial_\mu-{1\over 3}{g'^2\over G} Z_\mu+\tilde\nu X_\mu)d_R
\cr &
-G_d(\bar d_L\phi d_R +\bar d_R \phi^* d_L)
+G_u(\bar u_L\phi^* u_R + \bar u_R\phi u_L).
\cr}
\numeqn$$
where we have introduced couplings $\nu$, $\nu'$ and $\tilde \nu$ between the quarks and the $X$ field. 
Gauge invariance requires
$$
\tilde\nu=\nu+\alpha, \qquad \nu'=\nu-\alpha
\numeqn$$
where $\alpha$ is the charge of the electroweak Higgs field under the extra U(1) introduced in section 2. 
The generic form of the Lagrangian for each pair of particles is thus
$$
{\cal L}_{\rm f}=
\bar f_L \gamma^\mu(-i\partial_\mu -aZ_\mu +c X_\mu)f_L
+\bar f_R\gamma^\mu(-i\partial_\mu -bZ_\mu +d X_\mu)f_R
+m(\bar f_L\phi^* f_R + \bar f_R\phi f_L),
\numeqn$$
where $\phi$ should be replaced by $\phi^*$ in the case of the d quark and
electron. The Dirac equations  are then
$$
 \gamma^\mu(-i\partial_\mu -aZ_\mu +c X_\mu)f_L
+m\phi^* f_R=0
\numeqn$$
$$
\gamma^\mu(-i\partial_\mu -bZ_\mu +d X_\mu)f_R
+m\phi  f_L=0 
\numeqn$$
Using the Dirac matrices of ref.{\EP}
we have the following form for the left and right-handed spinors
$$
f_L=\pmatrix{f_1\cr f_2\cr -f_1\cr -f_2\cr}\qquad f_R=\pmatrix{f_3\cr f_4\cr 
f_3\cr f_4\cr}.
\numeqn$$

Making the ansatz
$$
\pmatrix{ f_1\cr f_2\cr f_3\cr f_4\cr}=\e^{iwt-ikz+in\theta}\pmatrix{A\cr B\e^{i\theta}\cr C\e^{il\theta}\cr D\e^{(l+1)i\theta}\cr},
\quad \phi=\e^{il\theta}\phi(r)
\numeqn$$
the Dirac equations reduce to
$$
\pmatrix{
w+k & -i({\partial_r+{n+1\over r}\atop -aZ_\theta+cX_\theta})&   m\vert\phi\vert   &  0 \cr
-i({-\partial_r +{n\over r}\atop -aZ_\theta+cX_\theta}) &     -w+k  & 0 & m\vert\phi\vert \cr
 -m\vert\phi\vert  &  0       & -w+k  & -i({-\partial_r-{n+l+1\over r}\atop+bZ_\theta-dX_\theta}) \cr
      0     & -m\vert\phi\vert  & -i({\partial_r-{n+l\over r} \atop+bZ_\theta-dX_\theta})    &w+k\cr
}
\pmatrix{ A\cr B\cr C\cr D\cr}=0
\numeqn$$

This is now in the form where the standard zero mode analysis [\JR] can now be 
applied. Let $w=k$ and set $A=D=0$, at small distances the leading
order terms in the Dirac equations reduce to
$$
-i(\partial_r+{n+1\over r})B + m\vert\phi\vert C=0,
\numeqn$$
$$
-m\vert\phi\vert B -i(\partial_r-{n+l\over r} ) C=0.
\numeqn$$
At small distances $\phi\propto r^{\vert l\vert}$ and we have two short distance solutions:
$$
\pmatrix{B \cr C\cr}\sim \pmatrix{ r^{n+l+\vert l\vert+1} +... \cr r^{n+l} +... \cr}
\numeqn$$
and
$$
\pmatrix{B \cr C\cr}\sim \pmatrix{ r^{-n-1} +... \cr r^{-n+\vert l\vert} +... \cr}.
\numeqn$$
Thus we have two regular small $r$ solutions for $n=-1,-2,..,-l$; we have $\vert l\vert$ zero modes for $l$ positive.

Similarly for $w=-k$ we set $B=C=0$ and the small distance leading
order terms in the Dirac equations reduce to
$$
-i(-\partial_r +{n\over r})A +m\vert\phi\vert D=0
\numeqn$$
and
$$
-i(-\partial_r-{n+l+1\over r}) D -m\vert\phi\vert A=0
\numeqn$$
which have small distances solutions,
$$
\pmatrix{A \cr D\cr}\sim \pmatrix{ r^{-n-l+\vert l\vert} +... \cr r^{-n-l-1} +... \cr}
\numeqn$$
and
$$
\pmatrix{A \cr D\cr}\sim \pmatrix{ r^{n} +... \cr r^{n+\vert l\vert+1} +... \cr}.
\numeqn$$
Both of these solutions are regular for $n=0,1,..,-l-1$; we have $\vert l\vert$ zero modes for $l$ negative.

The presence of the GUT string becomes important in the region between the GUT length scale and the fermion's length scale. Here the
GUT gauge field has attained its asymptotic form and the fermion mass term can still be neglected. The Dirac equations have a form
similar to those above with $n$ and $l$ shifted by contributions to the $1/r$ terms coming from the GUT gauge field.  
In this intermediate region some of the spinor components may vary as negative powers of $r$[\mandm]. However, this behaviour does not persist into the origin, such zero modes are not destroyed but are simply 'amplified' at the origin.

Hence we have shown that, even taking the core of the GUT string into account,
we have zero modes if $\vert \alpha/q\vert>1/2$. The existence of the zero
modes results in the string becoming superconducting at the electroweak
scale.

If the string does not wind there are no zero modes. However, the dip in the 
electroweak Higgs VEV constitutes a two-dimensional potential well.
Two-dimensional wells are similar to potential wells in one-dimension and
always have a bound state, unlike in three dimensions. Thus, there are always 
bound states in the Dirac spectrum.

Consider the ansatz above with $l=k=0$. The Dirac equation can then be written in the form 
$$
-\gamma^0\pmatrix{
 0& -i({\partial_r+{n+1\over r}\atop -aZ_\theta+cX_\theta})&   m\vert\phi\vert   &  0 \cr
-i({-\partial_r +{n\over r}\atop -aZ_\theta+cX_\theta}) &   0    & 0 & m\vert\phi\vert \cr
 -m\vert\phi\vert  &  0       & 0  & -i({-\partial_r-{n+1\over r}\atop+bZ_\theta-dX_\theta}) \cr
      0     & -m\vert\phi\vert  & -i({\partial_r-{n\over r} \atop+bZ_\theta-dX_\theta})    &0\cr
}
\pmatrix{ A\cr B\cr C\cr D\cr}=w\pmatrix{ A\cr B\cr C\cr D\cr}
\numeqn$$
and we can identify the Hamiltonian operator,
$$
\hat H=\pmatrix{
 0& i({\partial_r+{n+1\over r}\atop -aZ_\theta+cX_\theta})&   -m\vert\phi\vert   &  0 \cr
-i({-\partial_r +{n\over r}\atop -aZ_\theta+cX_\theta}) &   0    & 0 & m\vert\phi\vert \cr
 -m\vert\phi\vert  &  0       & 0  & -i({-\partial_r-{n+1\over r}\atop+bZ_\theta-dX_\theta}) \cr
      0     & m\vert\phi\vert  & i({\partial_r-{n\over r} \atop+bZ_\theta-dX_\theta})    &0\cr
}
\numeqn$$
If we can construct a state, $\psi$, such that $<\psi\vert \hat H\hat H\vert \psi>$ is less than $m^2\eta^2$ we can argue from the
variational principle that there is a state somewhere between the positive and negative energy continua.
Consider first the state
$$
\psi=\pmatrix{\zeta(r)\cr 0\cr0\cr0\cr},
\numeqn$$ 
then
$$
<\psi\vert \hat H\hat H\vert \psi>=\int d^2x \zeta^*\bigl(
m^2\vert\phi\vert^2+({\partial_r+{n+1\over r} -aZ_\theta+cX_\theta})
({-\partial_r +{n\over r} -aZ_\theta+cX_\theta})
\bigr)\zeta.
$$
$$
=2\pi\int rdr \bigl(
m^2\vert\phi\vert^2\vert\zeta\vert^2+\vert({-\partial_r +{n\over r} -aZ_\theta+cX_\theta})\zeta\vert^2
\bigr) +2\pi\bigl[r\zeta^*({-\partial_r +{n\over r} -aZ_\theta+cX_\theta})\zeta\bigr]_0^\infty,
\numeqn$$
after integrating by parts.
Now consider the profile function $\zeta$. Let $\zeta_0$ satisfy
$$
({-\partial_r +{n\over r} -aZ_\theta+cX_\theta})\zeta_0=0
\numeqn$$

The solution is  $\zeta_0\sim r^{ n}$ at small $r$ and, using the asymptotic forms of the gauge fields,
  $\zeta_0\sim r^{[n-a(2\alpha/Gq)+c(1/q)]}:=r^s$ 
at large $r$.  For $n=0$ the small $r$ form is a constant
and the large distance form is a decaying power law if the gauge field terms have the appropriate sign.
We now introduce an exponential damping factor to ensure that the state is normalisable 
and consider  $\zeta=\zeta_0 \e^{-\lambda r}$.

If this state is normalisable with $\lambda=0$ then only the first term  in (30) survives. The integral is less than the 
vacumm case when $\vert \phi\vert$ drops below its vacuum value. Alternatively, if a non-zero value of $\lambda$ is
required to produce a normalisable state, we have extra contributions to the expectation value, 
$$
<\psi\vert \hat H\hat H\vert \psi>=2\pi\int rdr \e^{-2\lambda r}\zeta_0\bigl(
m^2\vert\phi\vert^2  +\lambda^2 \bigr)\zeta_0 +2\pi\lambda\bigl[ \zeta_0^2(r\e^{-2\lambda r})\bigr]_0^\infty.
\numeqn$$

We are interested in small values of $\lambda$ and in this case the normalisation integral is dominated by the 
large distance contributions, giving  a normalisation factor proportional to $\lambda^{1+s}$. 
The dip in the Higgs field
occurs in the string core where the exponential factor can be
ignored. Thus the only $\lambda$ dependence in the mass term comes from the normalisation factor and hence
the energy reduction
due to the dip in the Higgs profile  varies as $\lambda^{2+2s}$. 
The boundary term in the expectatioin value vanishes and the extra contribution varies as  $\lambda^2$. 
  If $s$ is negative, 
by making $\lambda$ suitably small we can produce
a trial spinor whose squared energy lies below the continuum level. 

Similar arguments can be
constructed using a trial spinor with any one component nonzero. If the second component is taken to be nonzero, the 
order of the differential operators in (30) is reversed, the effective sign of the gauge field terms is reversed and this
component will provide a decaying $\zeta_0$ if the first component did not. 

We have shown that the string will have fermionic bound states if there is a dip in the scalar potential. These bound 
states can carry a current [\hillwidrow]. 

\bigskip
\noindent{\bf 5) Current Carrying Options}

We have constructed a composite string by considering couplings between 
the electroweak theory and a higher energy, string forming sector. We always
find a layer of nontrivial electroweak fields  surrounding the GUT string core. Depending on the 
ratio of gauge couplings, $\vert{\alpha/q}\vert$, the electroweak Higgs field 
can wind, leading to a stable electroweak string surrounding the GUT core.
This gives rise to fermion zero modes and the string becomes 
superconducting at the electroweak scale. The current carried will be of the
electroweak scale, this may be too small to produce dramatic cosmological
effects, but could alter some of the cosmological predictions
of the cosmic string scenario. This is under investigation.

For $\vert{\alpha/q}\vert < 1/2$ the electroweak Higgs field does not wind, 
there are no fermion zero modes,  but there are bound states which 
 can carry a current. In this case the current is smaller than in 
the zero mode case and is not persistent, the size and dissipation time 
being determined by the mass of the bound state[\hillwidrow].

\noindent{\bf Acknowledgements}

This work is supported in part by PPARC. We wish to thank M. Goodband
and A. Yates for discussions. 

\noindent{\bf References}
\item {1.} M.B. Hindmarsh and T.W.B. Kibble, Rep Prog Phys, {\bf 58} (1995) 
477; A. Vilenkin and E.P.S. Shellard, `Cosmic Strings and Other Topological
Defects' (Cambridge University Press) 1994  
\vskip 7pt
\ni
\item {2.} E. Witten, Nucl Phys {\bf 249} (1985) 557
\vskip 7pt
\ni
\item {3.} A.C. Davis and P. Peter, Phys Lett {\bf 358} (1995) 197
\vskip 7pt
\ni
\item {4.} R. Brandenberger, B. Carter, A.C. Davis and M. Trodden, hep-ph 9605382, 
\vskip 7pt
\ni
\item {5.} A. Everett, Phys Rev Lett {\bf 61} (1988) 1807; T.W.B. Kibble and 
A. Yates, private communication and in preparation; A. Yates, EU Conference, 
Capri (1996)
\vskip 7pt
\ni
\item {6.} C.T. Hill and L.M. Widrow, Phys Lett {\bf B189} (1987) 17;
M. Hindmarsh, Phys Lett {\bf B200} (1988) 429 
\vskip 7pt
\ni
\item {7.} R.L. Davis and E.P.S. Shellard, Nucl Phys {\bf B323} (1989) 209
\vskip 7pt
\ni
\item {8.} W.B. Perkins and A.C. Davis, Nucl Phys {\bf B406} (1993) 377
\vskip 7pt
\ni
\item{9}M. Goodband and M. Hindmarsh, Phys Lett {\bf B} (1996)
\vskip 7pt
\ni
\item {10.} A. Achucarro, R. Gregory, J. Harvery and K. Kuijken, Phys Rev Lett
{\bf 72} (1994) 3646
\vskip 7pt
\ni
\item {11.} M.A. Earnshaw and W.B. Perkins, Phys Lett {\bf B328} (1994) 337
\vskip 7pt
\ni
\item {12.} R. Jackiw and P. Rossi, Nucl Phys {\bf B190} (1981) 681
\end